\documentclass[12pt]{article}

\usepackage{amsmath}
\usepackage{amsfonts}
\usepackage{enumerate}
\usepackage{authblk}
\usepackage{mathtools, nccmath}
\usepackage{url} 
\usepackage{natbib}
\usepackage{xcolor}
\usepackage{adjustbox}
\usepackage{rotating}

\title{The appeal of the gamma family distribution to protect the confidentiality of contingency tables}

\author[$\dagger$]{James Jackson}
\author[$\star$]{Robin Mitra}
\author[$\dagger$]{Brian Francis}
\author[$\ddag$]{Iain Dove}
\affil[$\dagger$]{Lancaster University, Lancaster, UK}
\affil[$\star$]{University College London, London, UK}
\date{}
\affil[$\ddag$]{Office for National Statistics, Titchfield, UK}

\begin{document}
\maketitle
\begin{abstract}
Administrative databases, such as the English School Census (ESC), are rich sources of information that are potentially useful for researchers. For such data sources to be made available, however, strict guarantees of privacy would be required. To achieve this, synthetic data methods can be used. Such methods, when protecting the confidentiality of tabular data (contingency tables), often utilise the Poisson or Poisson-mixture distributions, such as the negative binomial (NBI). These distributions, however, are either equidispersed (in the case of the Poisson) or overdispersed (e.g.\ in the case of the NBI), which results in excessive noise being applied to large low-risk counts. This paper proposes the use of the (discretized) gamma family (GAF) distribution, which allows noise to be applied in a more bespoke fashion. Specifically, it allows less noise to be applied as cell counts become larger, providing an optimal balance in relation to the risk-utility trade-off. We illustrate the suitability of the GAF distribution on an administrative-type data set that is reminiscent of the ESC.
\end{abstract}

\bibliographystyle{rss}

\section{Introduction}
Large administrative databases, such as those held by government departments, are often rich sources of information, sometimes for entire populations. For this reason, the use of administrative data continues to be considered as an alternative source of data, despite the challenges involved \citep{Hand2018}. An example of a database that could be utilized for research purposes is the English School Census (ESC), held by the UK's Department for Education (DfE), which holds information about pupils attending state-funded schools in England. 

To protect the privacy of those individuals included, the dissemination of sensitive individual-level administrative records would necessitate  -- from both a legal and ethical perspective -- stringent application of statistical disclosure control (SDC) methods. Such methods perturb or suppress the original data until a satisfactory level of protection is achieved. The vast accumulation of literature attests to the amount of work conducted in SDC over the years; see, for example, \cite{Hundepool2012} and \cite{ Templ2017} and the references included therein. The use of synthetic data \citep{Rubin1993, Little1993, Drechsler2011} is a more recent addition to the SDC toolkit, the notion of which is to create new data by simulating from a model fit to the original data (synthesis model). In theory, synthetic data can preserve the statistical properties of the original data properties, while, as values are simulated, disclosure risk should be lower.  

The ESC -- as is the case more generally with administrative-type data -- is mainly composed of categorical variables, e.g.\ pupils' ethnicity, local authority, language, etc., and as such can be expressed as a contingency table, i.e.\ in aggregated format where counts give the frequencies with which the various combination of categories (cells) are observed. It is often intuitive to consider this tabular representation: firstly, from a confidentiality perspective it allows unique, i.e.\ at-risk, individuals to be identified, since these are those who belong to a cell with a count of one; secondly, from a modelling perspective, one needs only deal with univariate observations (counts), as opposed to multivariate observations.

When the ESC (or a portion of the ESC) is expressed as a contingency table, it is predominantly composed of zero counts. Yet there are also substantial numbers of small counts (e.g.\ in the range 1--10) that require protection, in addition to large counts (e.g.\ greater than 10) that are low-risk and hardly require protecting at all. This range of cell count sizes is much greater than is often seen in survey data sets because $n$, i.e.\ the ``sample size'', is large. 

Synthesis models for contingency tables broadly fall into one of two types: models that utilise multinomial-type distributions \citep{Abowd2008, Hu2014, Manrique-Vallier2018, Hu2018} and those that utilise count distributions, such as the Poisson or the NBI \citep{Graham2007, Jackson2021, Jackson2022, Quick2021,Quick2022}. Owing to the Poisson-multinomial equivalence \citep{Asmussen1983,Lang1996}, there are close links between the two types of approaches, but for the remainder of this paper we focus exclusively on the count-based approaches. 

A limitation of count-based approaches that use the Poisson (or Poisson-based) distribution(s), e.g.\ log-linear models, is that relatively more noise is applied to larger counts than to smaller ones, because the variance of such distributions increases as the mean increases. However, larger counts typically require less protection than smaller ones; it is the latter, particularly counts of one (uniques), that pose the greatest disclosure risk. To address this limitation, we propose the use of the discretized version of the gamma family (GAF) distribution (see \citealp{Rigby2019}), which is a three-parameter count distribution, where one of the parameters (later introduced as $\nu$) controls the mean-variance relationship. The use of the GAF simultaneously allows ``large'' counts to be modelled with an underdispersed distribution and ``small'' counts with an overdispersed one, resulting in relatively less noise being applied to larger counts than smaller counts.  

The remainder of the paper is organised as follows. Section \ref{sec2} describes limitations of existing Poisson-based approaches for synthesizing contingency tables. Section \ref{sec3} sets out the use of the (discretized) GAF distribution for synthesis. Section \ref{sec5} demonstrates the effectiveness of the GAF in relation to the negative binomial (NBI) when synthesizing the ESC$_\text{rep}$, a substitute database for the ESC. Section \ref{sec6} gives some concluding remarks.
\section{Approaches for synthesizing contingency tables and their limitations}\label{sec2}
Contingency tables comprise of a structured set of cell counts. We let $\mathbf{f}=(f_1, f_2, \hdots, f_K)$ denote the counts in the original data's contingency table (i.e.\ the \textit{original counts}) and $\mathbf{f}^\text{syn}=(f_1^\text{syn}, \hdots, f_K^\text{syn})$ denote the counts in the synthetic data's contingency table (i.e.\ the \textit{synthetic counts}), where $K$ denotes the number of cells in the table.
\subsection{Log-linear models} 
Synthesis models for contingency tables, as with the statistical analysis of contingency tables more generally, revolve around log-linear models (see \citealp{Agresti2013} for more about the analysis of contingency tables). The log-linear model can be expressed as a generalized linear model, in which it is parameterized by an intercept term, main effects and interaction effects. When using the Poisson log-linear model for synthesis \citep{Graham2007}, the synthetic counts are obtained by simulating from the following model:
\begin{align}
f_i^\text{syn} \mid \beta &\sim \text{Poisson}(\mu_i), \quad i=1,\hdots,K, \nonumber  \\
\text{where} \quad \text{log}(\mu_i) &= X_i\boldsymbol\beta. \label{poismodel}
\end{align}
and where $\beta\boldsymbol \in \mathbb{R}^d$ is the vector of log-linear model parameters and $X$ is the design matrix. The parameter vector $\boldsymbol\beta$ includes appropriate interactions; omitting, say, a two-way (and all higher order) interactions between a pair of variables is making an assumption that those variables are independent. Hence the synthesizer (i.e.\ he or she responsible for protecting the original data) must decide which interactions to include, and hence which relationships to preserve. To avoid this non-trivial modelling decision, in \cite{Jackson2021} we proposed the use of saturated log-linear models, which is the case where $\text{log}(\mu_i)= f_i$ in (\ref{poismodel}). We also proposed replacing the Poisson with a two-parameter count distribution, such as the negative binomial (NBI). If a random variable $X \sim$ NBI$(\mu , \sigma)$, then it has probability density function:
\begin{align*}
g_{N}(x ; \mu ,\sigma) &= \frac{\Gamma(x+{1}/{\sigma})}{\Gamma(x+1) \cdot \Gamma({1}/{\sigma})} \cdot \bigg(\frac{\sigma \mu}{1+\sigma \mu}\bigg)^{x} \cdot \bigg(\frac{1}{1+\sigma \mu}\bigg)^{1/\sigma}. 
\end{align*} When using the NBI in place of the Poisson, synthetic counts $\mathbf{f}^\text{syn}$ are obtained via the following mechanism:
\begin{align*}
f^\text{syn}_i \mid \mu_i, \sigma &\sim \text{NBI}(\mu_i, \sigma)   \\
\text{with} \quad \mu_i &= f_i. 
\end{align*}
The parameter $\sigma$, i.e.\ the NBI's scale parameter, can be used as a tuning parameter to control the level of noise applied to the original counts; increasing $\sigma$ would increase the noise. 

The use of the NBI distribution is analogous to the use of a Poisson-Gamma model \citep{Graham2007}. \cite{Quick2021} showed that such an approach can yield guarantees relating to differential privacy \citep{Dwork2006}.  
 \subsection{The limitation of Poisson-based distributions} \label{pois}
We henceforth group cell counts into two types: \textit{small} and \textit{large} counts. We define the former to be counts that are less than or equal to 10 and the latter to be those that are greater than 10. In SDC, counts that are roughly larger than around 10 are deemed to be safe, i.e. low risk of disclosure \citep{Willenborg1996,Ritchie2022}. Therefore, {small} counts can be considered \textit{at-risk} counts and {large} counts as \textit{safe} counts. We should perhaps further distinguish between \textit{small} counts and \textit{zero} counts, with the latter often warranting special consideration in their own right. 

A drawback when using Poisson or Poisson-based distributions is that, from a confidentiality perspective, noise is not necessarily applied in an optimal way. For a count of size $N$, the variance applied when using the Poisson is $N$ and when using the NBI it is $N + \sigma N^2$. The Poisson is equidispersed (variance equal to the mean) and the NBI is overdispersed (variance exceeds the mean). The reason that the NBI is overdispersed is that it is a continuously mixed Poisson distribution. To illustrate, suppose $X \sim$ Poisson($\lambda$) with mean and variance equal to $\lambda>0$. Now suppose that $X \mid \gamma \sim$ Poisson($\lambda \gamma$), where $\gamma$ is a continuous mixing distribution with mean equal to 1 and variance equal to $\sigma^2$. Then it follows that:
 \begin{align*}
     E(X) &= E[E(X\mid \gamma)] = \lambda \\
     \text{Var}(X) &= E[\text{Var}(X\mid \gamma)] + \text{Var}[E(X\mid \gamma)]  = E(\lambda \gamma) + \text{Var}(\lambda \gamma) = \lambda + \lambda^2 \sigma^2  > \lambda,
 \end{align*}
 hence mixing Poisson distributions adds variance, and results in overdispersed distributions.

As the variance of overdispersed (and equidispered) distributions increases as the size of the original count increases, relatively more noise is applied to large (low risk) counts than smaller (higher risk) ones. Instead, it would be preferable for small counts to be modelled by an overdispersed count distribution (to add appropriate noise) but large counts by an underdispersed ones (to add minimal noise). 

With this in mind, we can formulate the following research question: how can we generate synthetic counts using a single mechanism that applies noise in a more suitable way, i.e.\ applies far more noise to small counts than larger ones?

\section{The suitability of the discretized gamma family distribution (GAF)} \label{sec3}
Count distributions can be created by discretizing a continuous distribution defined on the interval $(0, \infty)$. In addition to the NBI, many other count distributions are overdispersed because of their origination as Poisson mixtures (e.g.\ the Poisson inverse-Gaussian, Delaporte, Sichel, etc.), but discretization can produce count distributions that are underdispersed. 

Let $W$ denote a continuous random variable defined on $(0, \infty)$, with probability density function (PDF) $f_W(w)$ and cumulative distribution function (CDF) $F_W(w)$. Then the corresponding discretised version of $W$, which we denote by $Y$, has probability mass function (PMF):
\begin{align}
f_Y(y) &=\begin{cases}
  F_W(1/2) & \text{if $y=0$ } \label{mech2a}   \\
     F_W(y + 1/2) - F_W(y-1/2) & \text{if $y = 1, 2, \hdots$}
  \end{cases}  
\end{align}
 \subsection{The gamma family distribution} \label{GAFdist}
\begin{figure}
\centering
        \includegraphics[height=6cm]{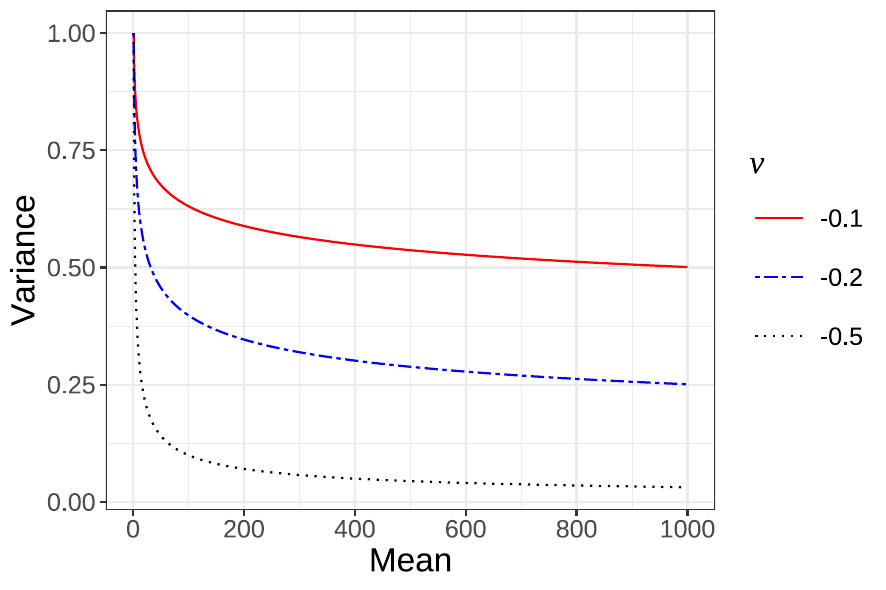}
\caption{The variance-mean relationship for three GAF distributions with different $\nu$ values (with $\sigma=1$). } 
    \label{GAF}
\end{figure}

 The gamma family distribution (see \citealp{Rigby2019}) is an extension to the gamma distribution. It has three parameters, $\mu>0$, $\sigma>0$ and $-\infty < \nu < \infty$ and its PDF and CDF are given as:
  \begin{align}
f_W(w \mid \mu, \sigma, \nu) &= \frac{w^{\sigma_1^{-2}-1}\text{exp}(-w\sigma_1^{-2} \mu^{-1})}{(\sigma_1^{2} \mu)^{\sigma_1^{-2}} \Gamma(\sigma_1^{-2})} \quad & \text{for}\quad y> 0 \nonumber \\
F_W(w \mid \mu, \sigma, \nu) &= \frac{\gamma(\sigma_1^2, w\mu^{-1}\sigma_1^{-2})}{\Gamma(\sigma_1^{-2})} \quad & \text{for}\quad y> 0 \nonumber
\intertext{{where} $\sigma_1=\sigma\mu^{\nu/2-1}$ and $\gamma(a,x)=\int_0^x t^{a-1}e^{-t} dt$ is the lower incomplete gamma function. Its mean and variance is:}
{E}[Y\mid \mu, \sigma ,\gamma] &= \mu \quad \text{and} \quad
\text{Var}[Y\mid \mu, \sigma ,\gamma] = \sigma^2 \mu^\nu. \label{GAF3}
\end{align}
The parameters $\mu$ and $\sigma$ have roughly similar roles to the corresponding $\mu$ and $\sigma$ in the NBI parameterization: $\mu$ controls the mean (location) and $\sigma$ (for a given $\mu$ and $\nu$) controls the variance (scale). The third parameter $\nu$ can be used to model the variance-mean relationship. When $\nu < 0$, the variance is a decreasing function of the mean. For a given $\mu$, the relationship between the variance and (negative) $\nu$ is one of exponential decay (or exponential growth if $\nu>0$). That is, the parameter $\nu$ controls the rate at which the variance falls away, i.e.\ the rate of decay. This is particularly useful in a synthesis context, as it allows us to control the rate at which the noise applied to larger counts falls away. Figure \ref{GAF} displays the variance-mean relationship for three GAF distributions with different $\nu$ values. For certain values of $\mu$, $\sigma$ and $\nu$, the GAF can be underdispersed. From (\ref{GAF3}), note that the GAF becomes underdispersed (mean less than the variance) when $\mu > \sigma^2\mu^\nu$, and rearranging gives that it becomes underdispersed when $\nu < -2\text{log}\;\sigma/\text{log}\;\mu$. 
\subsubsection{Comparison between the GAF and the Laplace}
The GAF is not unique as a continuous distribution that can be discretized to give an underdispersed count distribution. Another such example is the Laplace, which, of course, is widely used in differential privacy \citep{Dwork2006, Dwork2014}. When Laplace noise is added, constant noise is applied to cell counts, e.g.\ the same noise is applied when protecting an at-risk count of 1 as when protecting a safe count of 10. As the GAF can have a variance function that decreases as the mean increases, it has a potential advantage over the Laplace distribution.

 \subsection{Synthesis with the gamma family distribution (GAF)}
\begin{figure}
\centering
        \includegraphics[width=6cm]{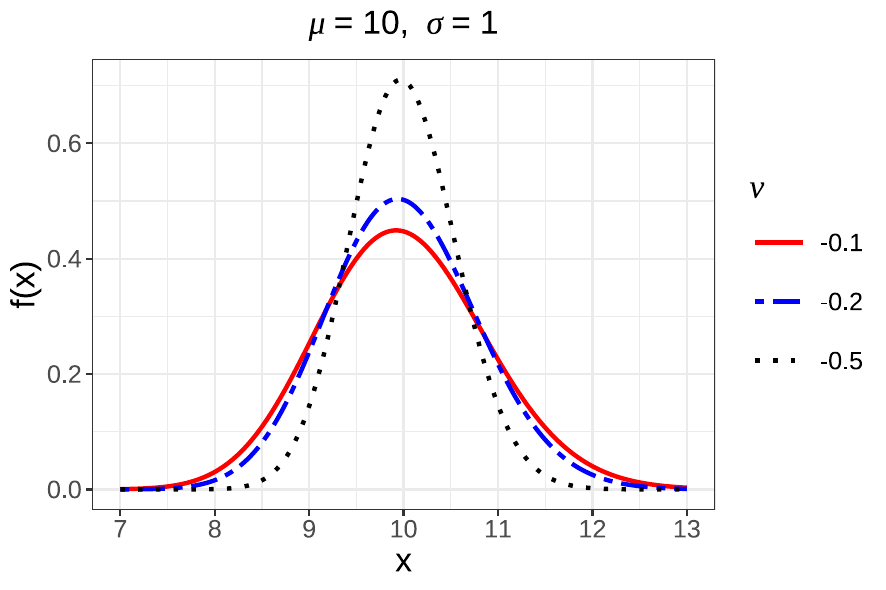}
          \includegraphics[width=6cm]{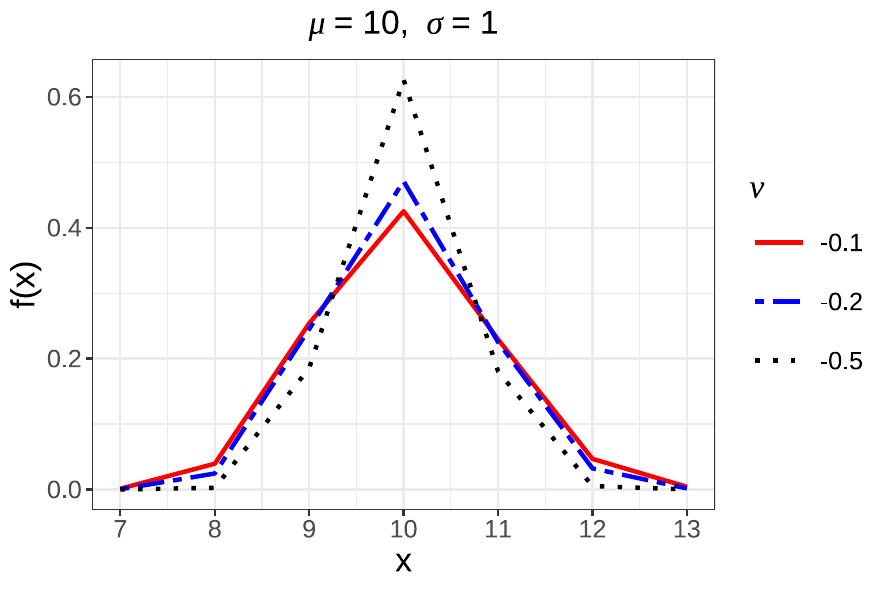}
\caption{The PDFs of the GAF (left plot) and GAF (right plot) distributions with $\mu=10$. } 
    \label{fig2}
\end{figure}

The GAF's mean and variance properties, given in (\ref{GAF3}), largely carry over to the discretized GAF (formed by applying the discretization procedure given in (\ref{mech2a})). Some small error is introduced because of the coarsening nature of discretization, but simulations have indicated that this is largely negligible. As such, the discretized GAF's variance function is roughly equal to $\sigma^2\mu^{\nu}$. For illustration, Figure \ref{fig2} shows the similarity between the GAF and GAF distributions when $\mu=10$ and $\sigma=1$. 

When using the GAF for synthesis, we therefore obtain synthetic counts $\mathbf{f}^\text{syn}$ by simulating from the following mechanism:
\begin{align}
f^\text{syn}_i \mid \mu_i, \sigma, \nu &\sim \text{GAF}(\mu_i, \sigma, \nu)  \quad i=1,\hdots,K, \nonumber \\
\text{with} \quad \mu_i &= f_i, \label{synmech3}
\end{align} 
and with values rounded to the nearest integer.
\subsection{Setting the tuning parameters} \label{sec3e}
For brevity, and to better illustrate the roles of $\sigma$ and $\nu$, we assume that saturated models are used, i.e.\ that $\mu_i=f_i$ in (\ref{synmech3}). This leaves us with two parameters, $\sigma$ and $\nu$, that can be used as tuning parameters. Both of these tuning parameters relate to variance and have a nice interpretability in a synthesis context. The quantity $\sigma^2$ can be thought of as the variance when $\mu=1$, i.e.\ the noise applied when synthesizing uniques (original counts of one). For a given $\sigma$, the parameter $\nu$ then controls the rate at which the variance falls away as the counts get larger, i.e.\ the extent to which noise is applied to large counts relative to small counts. 

 In practice, the actual setting of the tuning parameters is a policy decision and depends on the level of risk that the data protector deems acceptable. To aid this decision, the following can act as a rough guide to the setting of these parameters. Firstly, the synthesizer can first focus on small counts, and establish an acceptable level of protection for these. The parameter $\sigma$ is more effective for this, as for small counts (small $\mu$), the variance is largely dependent on $\sigma$, and less so on $\nu$. Secondly, the parameter $\nu$ can used for fine-tuning the noise applied to large counts, i.e.\ for controlling the rate at which the variance falls away. 
 
 The synthesizer can also use $\sigma$ and $\nu$ to ensure a particular value is obtained for a given risk or utility metric. This is discussed further in the next 
\subsection{Estimating risk and utility metrics \textit{a priori} with the GAF} \label{sec4}
Evaluating risk and utility is a key step when using any method for preserving confidentiality. When using saturated models with the GAF, risk and utility guarantees can be obtained \textit{a priori}, i.e.\ before any data protection actually takes places. We now give examples of metrics that can be estimated in this way by expressing them in terms of the GAF's tuning parameters, $\sigma$ and $\nu$.
\subsubsection{The $\tau$ metrics}
Let $f$ and $f^\text{syn}$ denote the original and synthetic count for an arbitrary cell. The $\tau$ metrics \citep{Jackson2021} are given as:
\begin{align*}
\tau_1(k)&=p(f^{\text{syn}}=k), & \tau_2(k)&=p(f=k)  \nonumber \\ \tau_3(k)&=p(f^{\text{syn}}=k | f=k) & \tau_4(k)&=p(f=k | f^{\text{syn}}=k). 
\end{align*}
The metrics $\tau_1(k)$ and $\tau_2(k)$ are the proportion of synthetic and original counts, respectively, of size $k$. The metric $\tau_3(k)$ is the probability that an original count of $k$ is synthesized to $k$; and the metric $\tau_4(k)$ is the probability that a synthetic count $k$ originated from a cell of size $k$. It is straightforward to see that $\tau_1(k)\tau_4(k)= \tau_2(k)\tau_3(k)$. 

When the GAF is used, the $\tau_3(1)$ metric, for example, can be expressed as functions of $\sigma$ and $\nu$:
\begin{align*}
\tau_3(k\mid \sigma, \nu) &=f_Y(y=k \mid \mu=k, \sigma, \nu) \\
&= \frac{\gamma(c^{-2}, (k+1/2)k^{-1}c^{-2})}{\Gamma(c^{-2})} - \frac{\gamma(c^{-2}, (k-1/2)k^{-1}c^{-2})}{\Gamma(c^{-2})} \quad \text{where} \quad c=\sigma k^{\nu/2-1} \\
&= \frac{\gamma \left[ \sigma^{-2} k^{-\nu
+2}, (k+1/2)k^{-1}(\sigma^{-2} k^{-\nu +2}) \right] }{\Gamma(\sigma^{-2}  k^{-\nu+2})} - \frac{\gamma \left[ \sigma^{-2} k^{-\nu+2}, (k-1/2)k^{-1}(\sigma^{-2} k^{-\nu +2}) \right]}{\Gamma(\sigma^{-2}  k^{-\nu+2})}.  
\end{align*} 
The infinite sum need not be calculated. Instead, it suffices to take the sum over the set of cell sizes observed in the original data, 
\begin{align}
    R=\{j \mid \tau_2(j)>0 \} \label{setR}.
\end{align}
Similar expressions can be found for the metrics $\tau_1(k)$ and $\tau_4(k)$. 
\subsubsection{Loss functions}
Loss functions tend to be viewed as measures of utility because small loss tends to imply high utility. Yet they are equally relevant to risk because small loss also implies reduced confidentiality. 

The \textit{quadratic loss function} (\textit{squared error}) is arguably the most well-known; its expectation (known as the \textit{mean squared error}) is given as:
 \begin{align*}
L_1 &= E \left[  \sum\limits_{k=1}^K \left( f_k-f_k^\text{syn} \right) ^2 \right] = \sum\limits_{k=1}^K   \Big[\text{Var}\left(f_k^\text{syn} \right) \Big]. 
\intertext{Substituting the GAF's approximate variance function (given in \ref{GAF3}) leads to:}
  L_1(\sigma, \nu, m) &\approx \sum\limits_{k=1}^K   \sigma^2 f_k^{\nu}/m  =  \sum\limits_{j>0, j \in R}  \left[ K \cdot \tau_2(j) \cdot  \sigma^2 j^{\nu}/m \right],
\end{align*}
where $R$ is as defined in (\ref{setR}). Original counts of zero ($f_k=0$), can be ignored as $L_1$ is undefined for zero counts when $\nu<0$. 
\subsubsection{Marginal counts}
When using this GAF mechanism, all marginal counts in the synthetic data's contingency table are unbiased. The largest marginal count is the ``grand total'', i.e.\ $n_\text{syn}=\sum_{i=1}^K f_i^\text{syn}$, the sum of {all} $K$ synthetic counts. We can estimate the variance of $n_\text{syn}$:
\begin{align*}
\text{Var}(n_\text{syn})&= \text{Var} \left( \sum\limits_{k=1}^K f_k^\text{syn} \right) = \sum\limits_{k=1}^K \text{Var} \left(  f_k^\text{syn} \right)  \approx \sum\limits_{k=1}^K  \sigma^2  f_k^{\nu},
\end{align*} which is identical to the expression derived for $L_1$. In large tables (large $K$), the central limit theorem establishes that $n_\text{syn}$ is normally distributed with this mean and variance. This can be used, say, to calculate the probability that $n_\text{syn}$ will be within $d$ of $n$, which may be useful in population data where $n$ is known:
\begin{align*}
p(|n_\text{syn}-n| < d) & = p(n_\text{syn} < n+ d)-p(n_\text{syn} < n- d) \\
& \approx \Phi \left[\frac {(n+d)-n }{\sigma \left( \sum_{k=1}^K  f_k^{\nu}\right)^\frac{1}{2} } \right]-\Phi \left[\frac {(n-d)-n }{\sigma \left( \sum_{k=1}^K  f_k^{\nu}\right)^\frac{1}{2} } \right] =2\Phi \left[\frac {d }{\sigma \left( \sum_{k=1}^K  f_k^{\nu}\right)^\frac{1}{2} } \right].
\end{align*}
Similar expressions can be obtained for any marginal count by summing over the relevant subset of counts. For those count distributions such as the Poisson that have identities for sums of random variables, marginal counts' exact distributions can be found. Nevertheless, when $K$ is large, normal approximations should suffice.
\section{An empirical demonstration of the GAF} \label{sec5}
\subsection{The data}
The actual ESC data is not currently available to researchers, even for the sake of demonstrating the effectiveness of methods for preserving confidentiality. Instead, we use a representative data set, ESC$_\text{rep}$, which was created by the Office for National Statistics (ONS) using only publicly available data sources, such as published data from the ESC and the 2011 UK Census data (see \citealt{Blanchard2022}). Importantly, the structure of the ESC$_\text{rep}$ is reminiscent of a subset of the ESC, and therefore gives an accurate reflection of performance on the ESC. 

When expressed as a contingency table,the ESC$_\text{rep}$data has approximately $3.5 \times 10^6$ cells, around 90$\%$ of which are zeros (with no structural zeros). Table \ref{datasummary2} provides a list of the variables and their numbers of categories; while Table \ref{datasummary3} looks at the cell count sizes, showing that there are indeed substantial numbers of both small and large counts. 
   \begin{table}
\caption{\label{datasummary2} A summary of the variables in the ESC$_\text{rep}$ data set \citep{Blanchard2022}.}
\centering
\begin{tabular}{*{3}{l}}
Name & Description & No. of Categories \\
\hline 
GEOGRAPHY & Pupil’s local authority & 326 \\
ETHNICITY & Pupil’s ethnicity & 20 \\
SEX  & Pupil’s sex & 4 \\
AGE  & Pupil’s age & 19 \\
LANGUAGE & Pupil's first language & 7 \\
\end{tabular}
\end{table} 
\begin{table}
\caption{\label{datasummary3} A summary of the cell sizes in the ESC$_\text{rep}$ data set.}
\centering
\begin{tabular}{*{3}{c}}
Cell count & Frequency & $\%$ of cells \\
\hline 
0 & 3,134,980 & 90.38 \\
1 & 119,917 & 3.46 \\ 
2 & 51,412 & 1.48 \\ 
3 & 25,952 & 0.75 \\
4 & 19,450 & 0.56 \\
5 & 13,076 & 0.38 \\
6 & 10,345 & 0.30  \\
7 & 7,947 & 0.23 \\
8 & 7,077 & 0.20   \\
9 &  5,809 & 0.17  \\
10 & 5,163 & 0.15 \\
$11 \leq$ & 67,512 & 1.95  \\  \hline
Total & 3,468,640 & 100
\end{tabular}
\end{table}
\subsection{Generating the synthetic data sets}
We synthesize the data using the GAF distribution and, for comparison, the NBI distribution, considering a range of parameter values. We consider saturated models to clearly illustrate the role of $\sigma$ and $\nu$ as tuning parameters (for example, fitting an all two-way log-linear model would introduce error from model uncertainty and it would be difficult to distinguish this uncertainty from the uncertainty introduced by the tuning parameters). As saturated models are used, we add a psuedocount of $\alpha=0.01$ to the zero counts (see \citealp{Jackson2021}) to open the possibility that zeros are synthesized to non-zeros. 

For the GAF, we use nine combinations of parameter values that relate to the cases involving $\sigma=0.5$, 1 and 2 and $\nu=0$, -0.25 and -0.5. For the NBI, we use $\sigma=0.5$, 1 and 2. As is common in the syntthetic data literature, we generate multiple -- specifically, $m=10$ -- synthetic data sets, which allows combining rules to be applied and valid inferences to be obtained  \citep{Raghunathan2003, Reiter2003, Reiter2005b, Drechsler2011}. The functions \texttt{rGAF} and \texttt{rNBI} from the {R} package \textbf{gamlss.dist} \citep{Stasinopoulos2007, Stasinopoulos2022} were used to generate GAF and NBI synthetic counts, respectively.
\subsection{The results}
\begin{figure}
\centering
        \includegraphics[width=6cm]{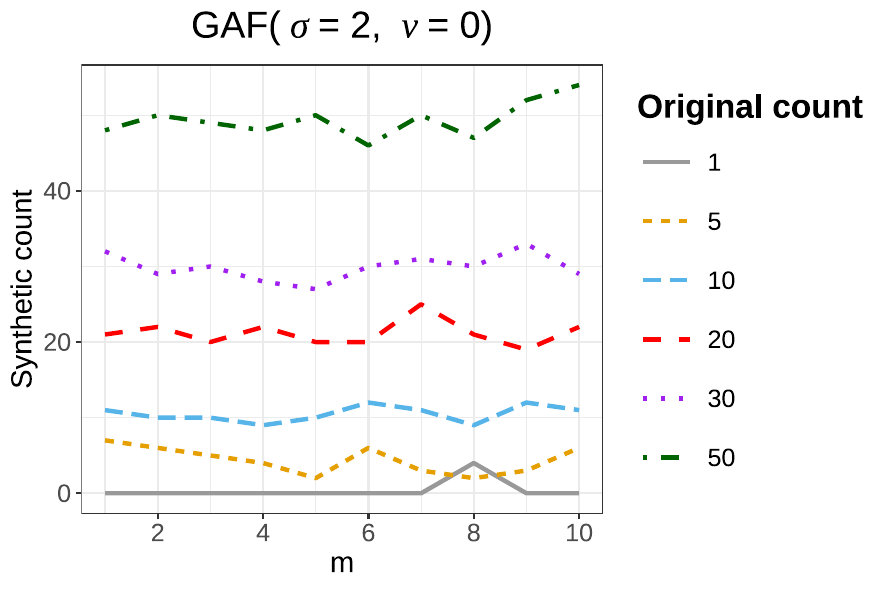}
            \includegraphics[width=6cm]{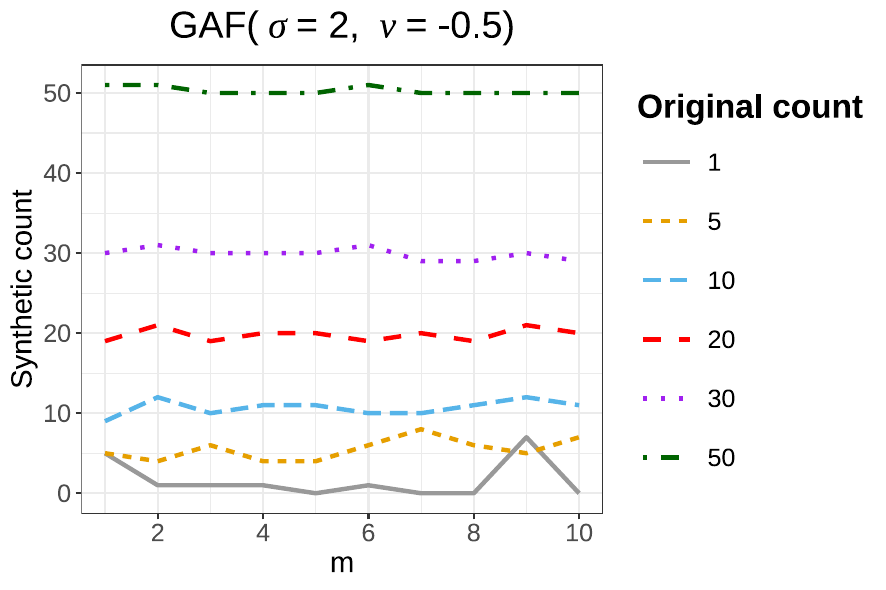}
            \includegraphics[width=6cm]{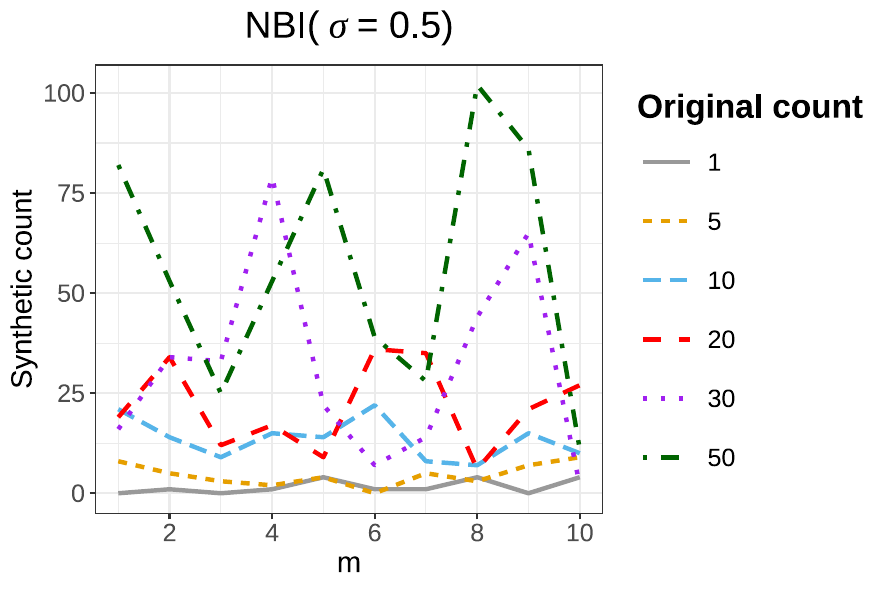}
\caption{For a selection of original counts, $m=10$ corresponding synthetic counts. The top plots relate to the GAF and the lower plot relates to the NBI.} 
    \label{syncounts}
\end{figure}

Figure \ref{syncounts} gives examples of $m=10$ synthetic counts obtained for original counts of different sizes. The top two plots relate to the GAF and the bottom plot to the NBI. While similar amounts of noise are applied to small counts (i.e.\ count of 1 and 5, denoted by the grey and yellow lines, respectively), the NBI applies far more noise to the larger counts (e.g. to the original count of 50, denoted by the green line). The two GAF plots also help to illuminate the role of the parameter $\nu$: when $\nu=-0.5$ (top right plot) similar noise is applied to smaller counts than when $\nu=0$, but less noise is applied to large counts.  

\begin{figure}
\centering
        \includegraphics[height=4.5cm]{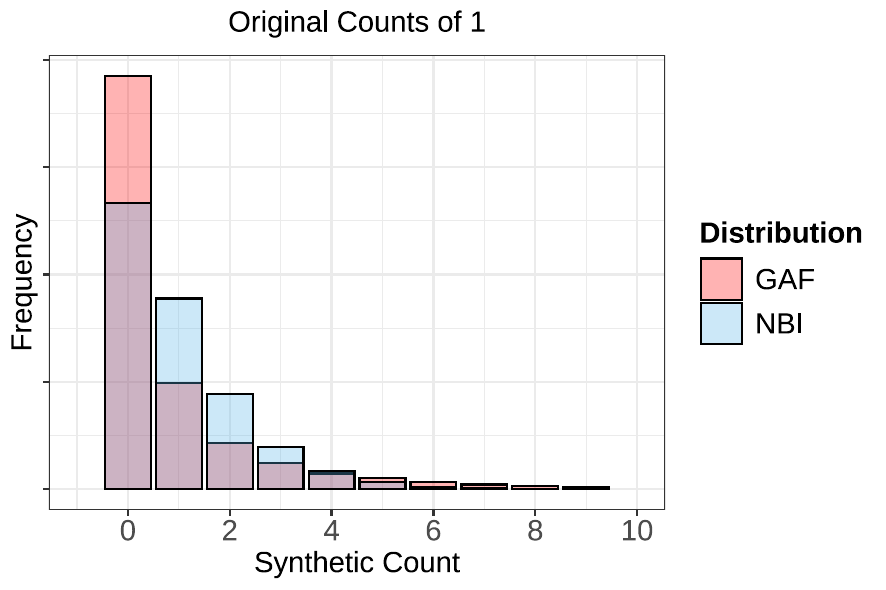}
        \includegraphics[height=4.5cm]{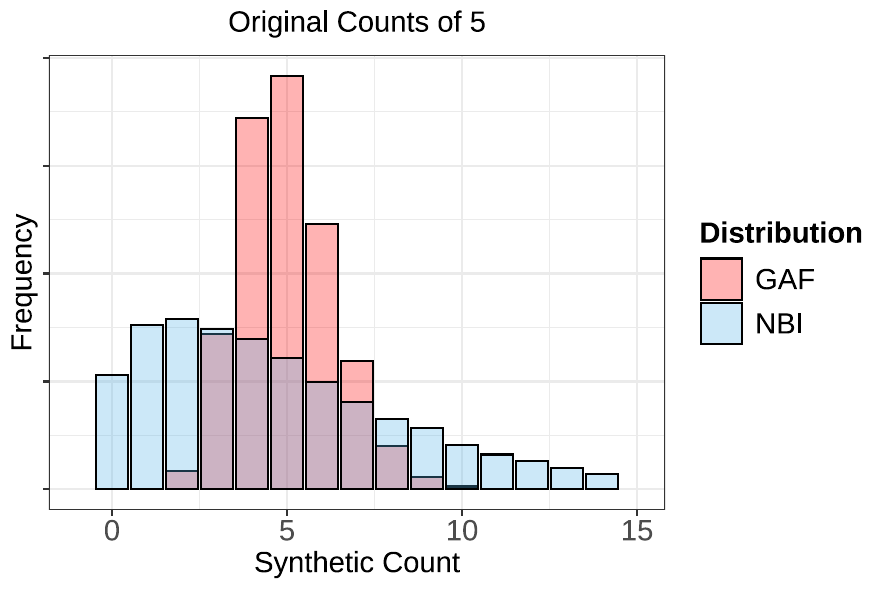}
        \includegraphics[height=4.5cm]{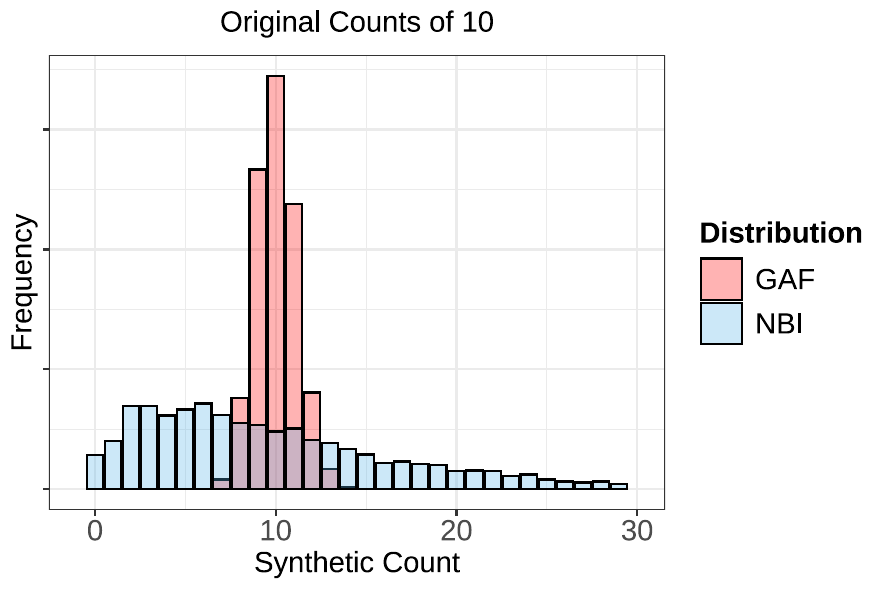}
        \includegraphics[height=4.5cm]{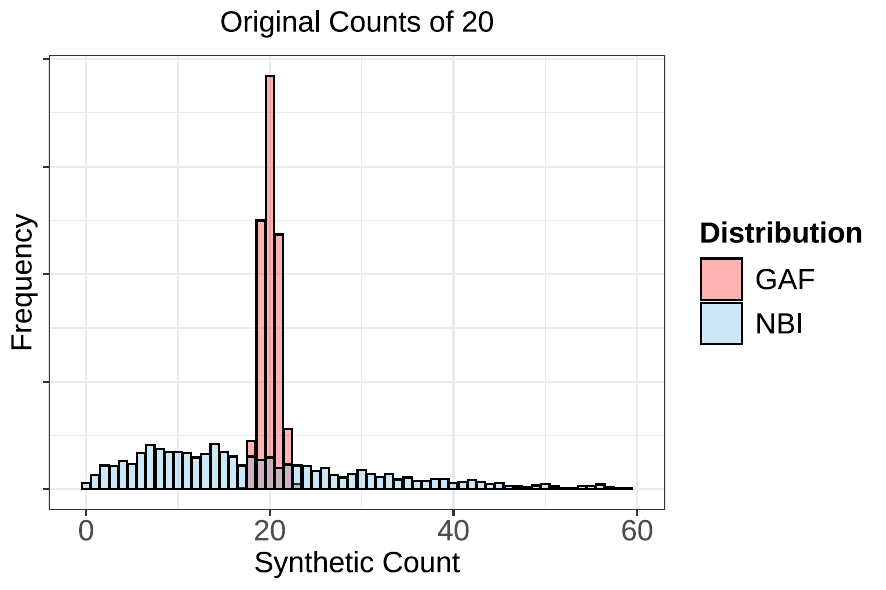}
\caption{The range of synthetic counts obtained for original counts of 1 (top-left), 5 (top-right), 10 (bottom-left) and 20 (bottom-right), for the GAF with $\sigma=2$ and $\nu=-0.5$ and for the NBI with $\sigma=2$.}
    \label{fig14}
\end{figure}

The bar charts in Figure \ref{fig14} compare the range of synthetic counts obtained from various sizes of original count; specifically, we look at original counts of 1 (top-left), 5 (top-right), 10 (bottom-left) and 20 (bottom-right). We compare the GAF with $\sigma=2$ and $\nu=-0.5$ to the NBI with $\sigma=0.5$. The top-left plot of Figure \ref{fig14} shows the range of synthetic counts obtained from original counts of one; this is essentially the empirical value of the metric $\tau_3(1)$ (i.e.\ the proportion of ones that were synthesized to one), and can thus be viewed as a risk measure. In this instance, the risk for the GAF and NBI are roughly similar. The other bar charts in Figure \ref{fig14} are perhaps better viewed as measures of utility, since these focus on larger (and hence lower risk) counts. In each of these the GAF produces synthetic counts that are much closer to the original counts. The bars in Figure \ref{fig14} would always be roughly the same regardless of the data being synthesized; as the number of cells increases, these proportions would tend towards their true values, which can be derived analytically, as shown in Section \ref{sec4}.  \par
\begin{figure}
\centering
        \includegraphics[height=4.5cm]{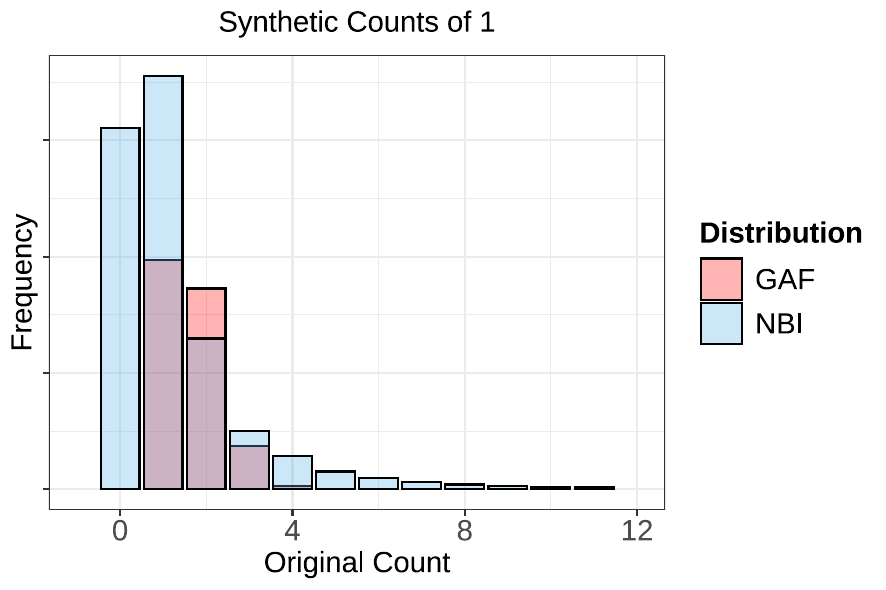}
        \includegraphics[height=4.5cm]{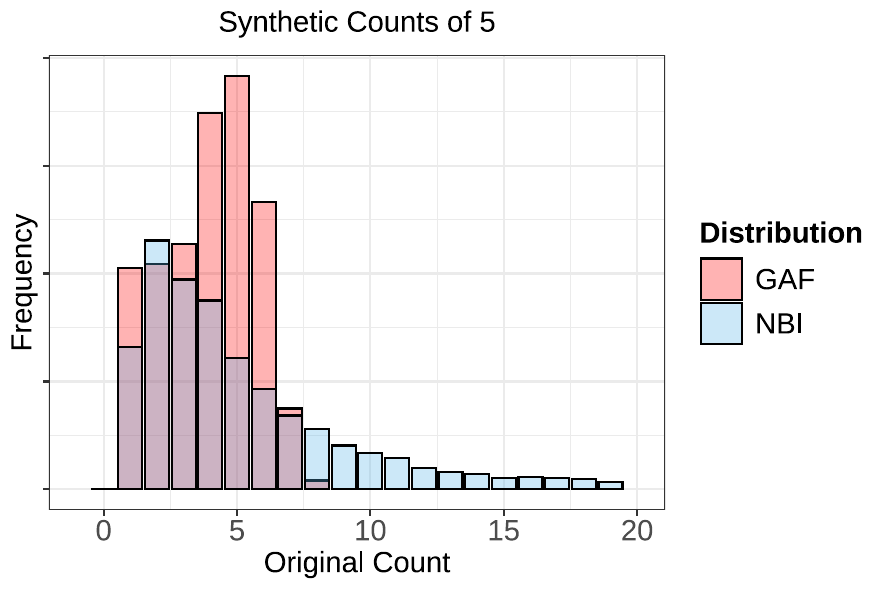}
        \includegraphics[height=4.5cm]{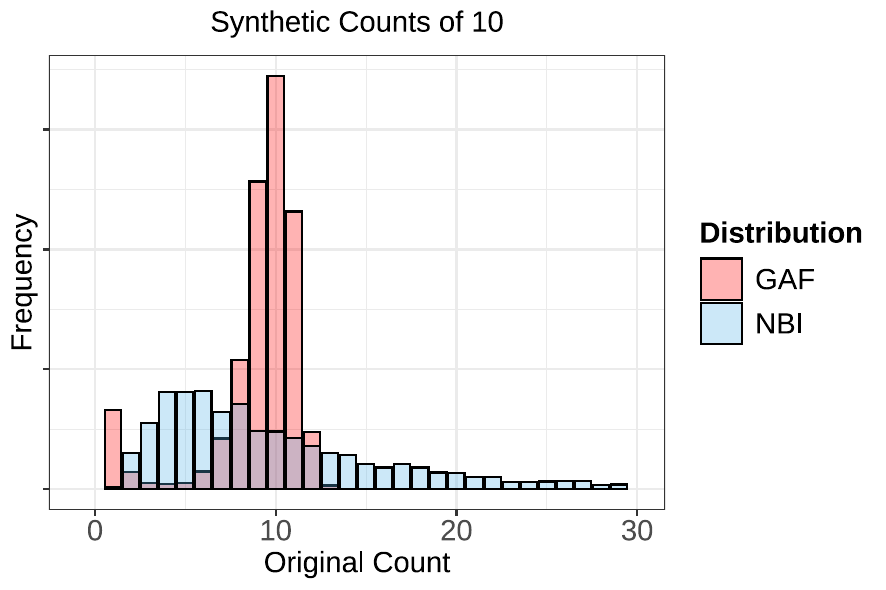}
        \includegraphics[height=4.5cm]{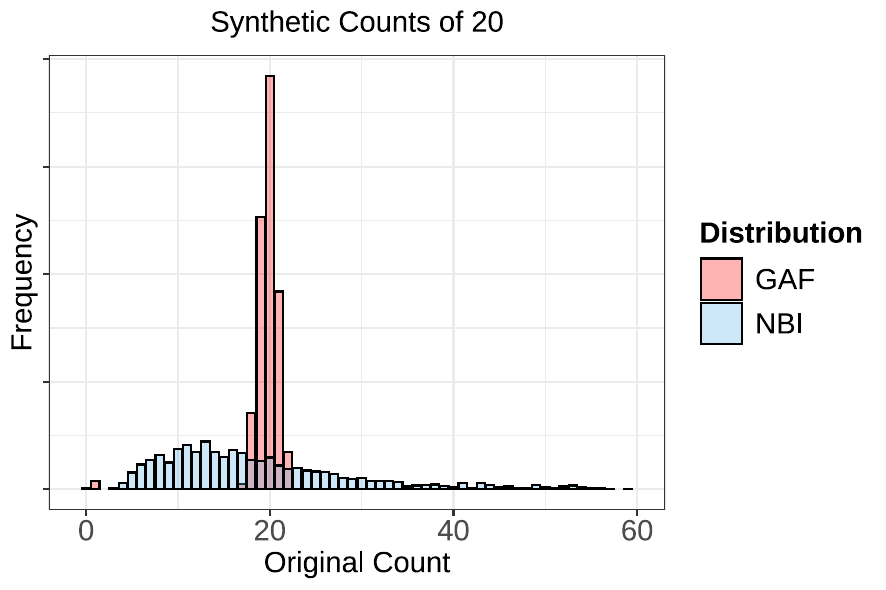}
\caption{The range of original counts relating to synthetic counts of 1 (top-left), 5 (top-right), 10 (bottom-left) and 20 (bottom-right), for the GAF with $\sigma=2$ and $\nu=-0.5$}
    \label{fig15}
\end{figure}

 The bar charts in Figure \ref{fig15} compare the range of original counts relating to various sizes of synthetic count when using the GAF with $\sigma=2$ and $\nu=-0.5$ and the NBI with $\sigma=2$; we look again at counts of 1 (top-left), 5 (top-right), 10 (bottom-left) and 20 (bottom-right). The larger synthetic counts (e.g.\ 10 and 20) tended to originate from original counts of similar sizes. The top-left plot of Figure \ref{fig15} shows the range of original counts relating to synthetic counts of one. This highlights one of the limitations of the GAF: that no synthetic counts of one originated from original counts of zero. Specifically, this highlights the failure of additive smoothing (adding small $\alpha$ to zero counts) to convert zeros into non-zeros. This is particularly a problem when saturated models are used. When using the NBI, adding $\alpha=0.01$, roughly equates to 1 in 100 counts being synthesized to a 1. With the GAF, however, when $\mu$ is small and $\nu < 0$, the variance becomes very large and, as such, adding $\alpha$ does not have the desired effect. Alternative ways to convert zeros to non-zeros will be required, such as randomly converting a proportion of zeros to ones using a Bernoulli mechanism. 
\begin{figure}
\centering
        \includegraphics[width=4cm]{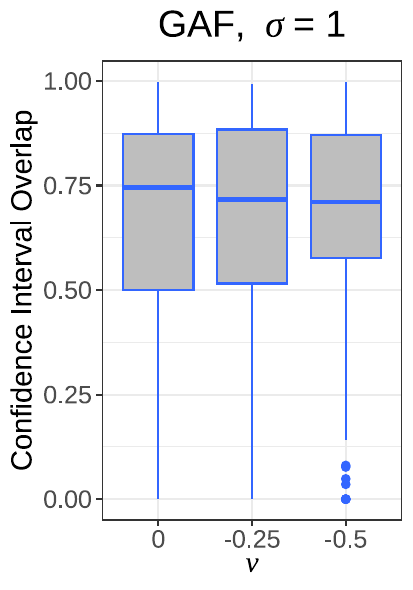}
        \includegraphics[width=4cm]{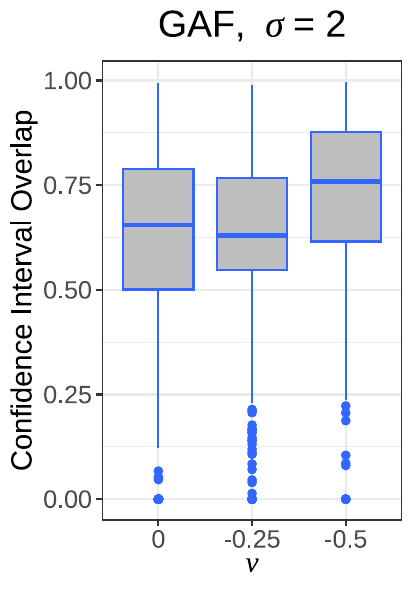}
        \includegraphics[width=4cm]{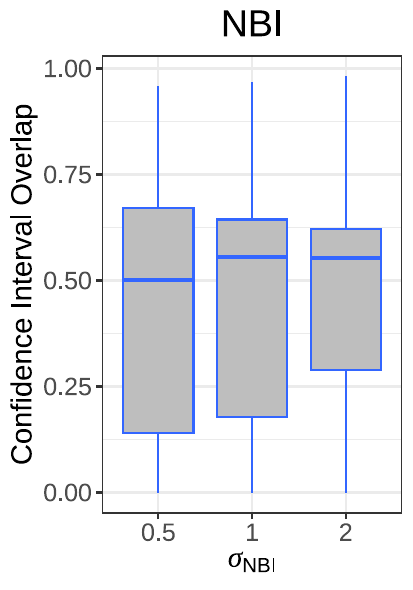}
\caption{The parameter estimates' overlap for a model fit to the original and synthetic ESC$_\text{rep}$ data sets. The left and middle plots are for the GAF and the right plot is for the NBI.}
    \label{fig17}
\end{figure}
 \subsection{Evaluating specific utility}
To evaluate specific utility \citep{Snoke2018}, we also fit a model to the original and synthetic ESC$_\text{rep}$ data sets; specifically, we fit an all-two-way interaction log-linear model to the ETHNICITY, AGE and LANGUAGE variables. \par  The confidence interval overlap for the model's parameter estimates are given as boxplots in Figure \ref{fig17}. The median overlap, for example, when using the GAF is higher than that of the NBI. Interestingly, altering the GAF or NBI's tuning parameters appears to have little effect on the median overlap. 

\subsection{Combining risk and utility}
\begin{figure}
\centering
\includegraphics[height = 7cm]{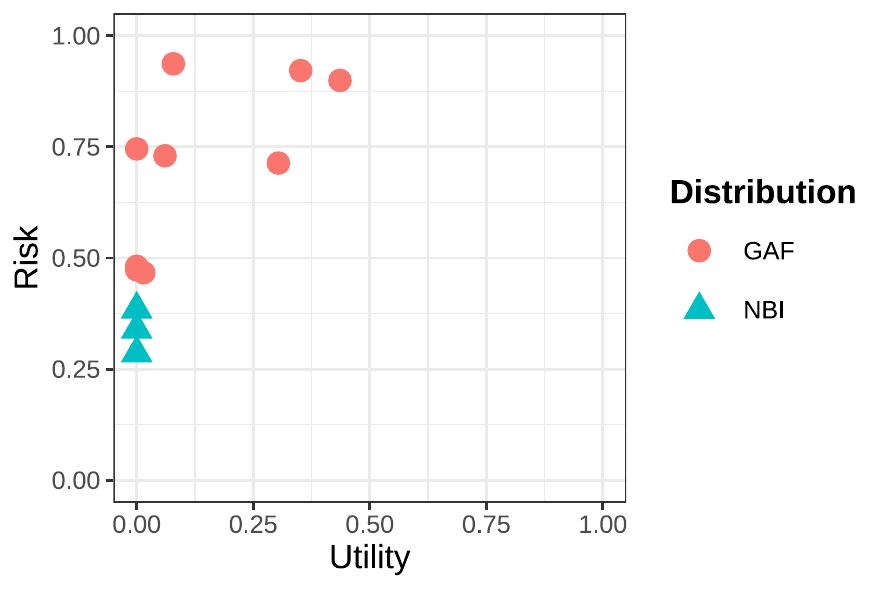}
\caption{\label{fig08} A risk-utility trade-off plot. Risk is measured by the $\tau_4(1)$ metric and utility by the mean squared error. The optimal position (lowest risk and highest utility) is the point (0,1).}
\end{figure}

Figure \ref{fig08} gives an example of a risk-utility trade-off (see \citealp{Duncan2001}), where risk, measured by the $\tau_4(1)$ metric, is plotted against utility, which is measured by the mean squared error. To allow risk and utility to be measured on a unit square, we take the inverse-logit of the mean squared error and subtract this value from 1, such that 1 represents high utility and 0 low utility. Points have been plotted for the various $\sigma$ and $\nu$ values considered. Points relating to the GAF (denoted by circles), are higher risk than the NBI, but also provide higher utility, hence demonstrating the role that the risk-utility trade-off plays when deciding on suitable synthesis methods. 
\section{Discussion} \label{sec6}
 The gains in utility when using the GAF stem from its ability to model the mean-variance relationship and hence gradually switch from an overdispersed distribution to an underdispersed one as counts become larger. Although we have illustrated the GAF when using saturated models, the appeal of the GAF would carry over to when, for example, an all two-way interaction log-linear model is used, because less noise would still be applied to large counts than small counts. Another appealing feature of the GAF is the intuitiveness of its tuning parameters, with $\sigma$ roughly relating to the variance applied to small counts and $\nu$ to the rate at which variance falls away. As shown in Section \ref{sec4}, these parameters can also be set according to a metric such as a loss function. One area that requires further consideration, though, is the use of additive smoothing in relation to the GAF, especially when using saturated models. 
 
 Data confidentiality methods are often, somewhat unfortunately, pigeonholed into various types; they may, for example, be pigeonholed as SDC methods or synthetic data methods or differential privacy (DP) methods. By treating these methods as separate entities, appealing aspects are lost. As an example, DP methods hardly ever consider generating multiple data sets, as is commonly done in synthetic data, as a way to obtain valid inferences. Another example is that SDC methods often make the distinction between sensitive and safe counts, with only sensitive counts requiring perturbation. This distinction is not made in DP, however: all counts are perturbed by the same amount of noise. With the GAF, there may be scope to develop a framework that has DP's ability to obtain formal risk guarantees, while also having the traditional SDC philosophy of sensitive and safe counts. 
\bibliography{main}       
\end{document}